%% file: paper.tex
\documentclass[conference,a4paper]{IEEEtran}
\pdfoutput=1

\usepackage[T1]{fontenc}
\usepackage[utf8]{inputenc}
\usepackage{balance}
\usepackage{amsmath,amssymb,amsfonts,yhmath,mathrsfs}
\usepackage{cite}
\ifCLASSINFOpdf
  \usepackage[pdftex]{graphicx}
\else
  \usepackage[dvips]{graphicx}
\fi
\usepackage{orcidlink}
\usepackage{url}
\usepackage{xcolor}
\usepackage{multirow}

\ifCLASSOPTIONcompsoc
 \usepackage[caption=false,font=normalsize,labelfont=sf,textfont=sf]{subfig}
\else
 \usepackage[caption=false,font=footnotesize]{subfig}
\fi

\usepackage{bm}

\usepackage{hyperref}

\title{{\fontsize{23.8}{33.8}\selectfont Bistatic Reflectivity and Micro-Doppler Signatures of Drones for Integrated Communication and Sensing}}

\author{%
\IEEEauthorblockN{%
Heraldo Cesar Alves Costa\IEEEauthorrefmark{1}\,\orcidlink{0009-0002-6186-5780},
Saw James Myint\IEEEauthorrefmark{1}\,\orcidlink{0009-0007-3788-7126},
Carsten Andrich\IEEEauthorrefmark{1}\,\orcidlink{0000-0002-4795-3517},
Sebastian W. Giehl\IEEEauthorrefmark{1}\,\orcidlink{0009-0008-1672-1351}, \\
Christian Schneider\IEEEauthorrefmark{1}\IEEEauthorrefmark{2}\,\orcidlink{0000-0003-1833-4562},
Reiner S. Thom\"a\IEEEauthorrefmark{1}\,\orcidlink{0000-0002-9254-814X}
}
\IEEEauthorblockA{\IEEEauthorrefmark{1}
Institute for Information Technology and Thuringian Center of Innovation in Mobility,\\
Technische Universität Ilmenau, Ilmenau, Germany}
\IEEEauthorblockA{\IEEEauthorrefmark{2}Fraunhofer Institute for Integrated Circuits IIS, Ilmenau, Germany}
{heraldo-cesar.alves-costa, saw-james.myint@tu-ilmenau.de}
}

\begin{document}
\maketitle

\begin{abstract}


The integration of wireless communication and radar sensing is gaining the interest of researchers from wireless communication and radar societies.
Sensing in Integrated Communication and Sensing (ICAS) systems differs from the traditional radar system in the configuration of transmitter-target-receiver, the operating frequency bands, and the transmitting waveform.
It is necessary to understand how target electromagnetic signatures behave in this context.
Therefore, this paper presents measurements and analysis of two important target signatures, reflectivity and micro-Doppler, for sensing in ICAS.
These target signatures are measured in the state-of-the-art measurement system, \underline{Bi}statische-\underline{Ra}dar-Messeinrichtung (BiRa).

\end{abstract}

\vskip0.5\baselineskip

\vspace{5pt}
\section{Introduction}\label{Introduction}
\input{1Introduction.tex}

\vspace{5pt}
\section{Static Reflectivity Signature}\label{Reflectivity}
\input{2Reflectivityv2.tex}


\vspace{5pt}
\section{Micro-Doppler Signature}\label{Micro-Doppler}
\input{3Micro-Doppler.tex}


\vspace{5pt}
\section{Conclusions and Outlook}\label{Conclusion}
\input{4Conclusions.tex}

\vspace{5pt}
\section*{Acknowledgment}
The authors would like to thank Dr.Ing. Tobias Nowack and M.Sc. Masoumeh Pourjafarian for their support throughout the BiRa measurement in ThiMo.
This research is funded by the BMBF project 6G-ICAS4Mobility, Project No.16KISK241, by the Federal State of Thuringia, Germany, and by the European Social Fund (ESF) under grants 2017 FGI 0007 (project "BiRa") and 2021 FGI 0007 (project "Kreat\"or").
\vspace{5pt}
\bibliographystyle{IEEEtran}
\bibliography{references}

\end{document}

%% file: 1Introduction.tex
With the growth in importance of green and re-usability technologies, the decades-long independently developing systems, mobile radio, and radar sensing are now converging in recent years \cite{Thoma0}, \cite{Thoma1}. 
With the advancements in software-defined radio (SDR), computation, and Software in the Loop technologies, the resource requirement and implementation gap between mobile radio and radar systems become narrower.
As the use of radar sensing is increasing, and emerging of the cognitive radar system, it needs resource allocation schemes and network control like mobile communication.
On the other hand, with the fast-growing connectivity of devices, mobile communication needs more frequency bands, and sharing with radar sensing is one of the solutions.
Therefore, the integration of the two systems becomes popular within a few years. Even though many variants of names and ideas can be found in literature, it can be categorized in general as 1. communication-centric integration 2. sensing-centric integration, and 3. developing both as a single system.  
At the moment we are focusing on the communication-centric system, which is known as the 
Integrated Communication and Sensing (ICAS).

\begin{figure}[b]
    \centering
    \includegraphics[width=3.0in, height=2.0in]{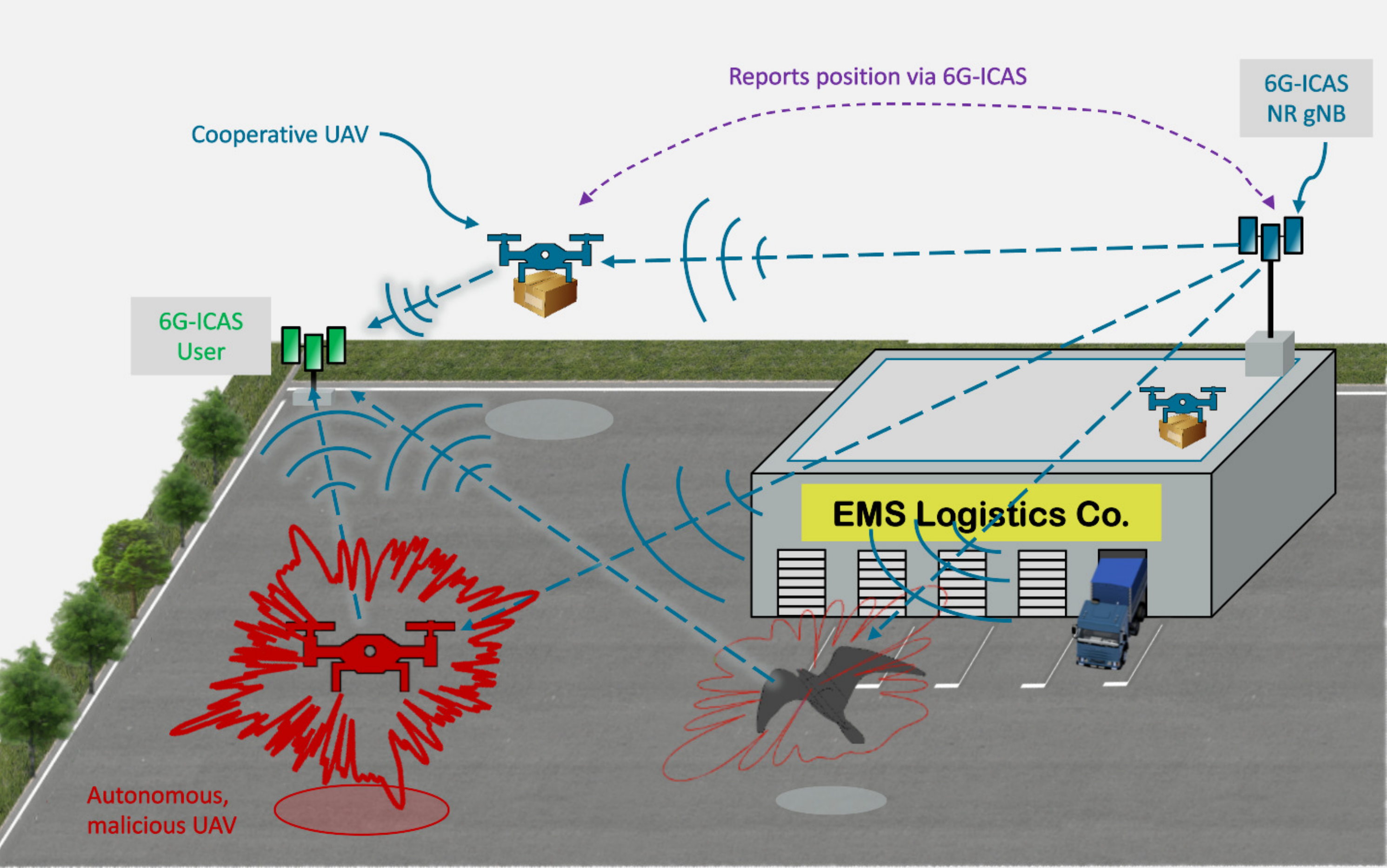}
    \caption{An example of ICAS surveillance scenario}
    \label{ICAS_Surveillance_Scenario}
\end{figure}

In \autoref{ICAS_Surveillance_Scenario}, an example of a communication-centric version of the ICAS surveillance scenario is illustrated. 
In this logistic center scenario, the cooperative delivery drones periodically report their cooperative information, such as identification number, position, trajectory, etc. to the base station (6G-ICAS NRgNB). 
At the same time, the passive sensor (6G-ICAS user) and the base station are able to detect (monostatically or bistatically) the positions and trajectories of the cooperative delivery drones due to the communication signals between the base station and the cooperative drones. 
Then the detection and cooperative information are shared (or exchanged) between the passive sensor and the base station. 
The mismatch between exchanged information indicates non-cooperative behavior within the surveillance area, and therefore, the necessary actions can be activated to detect, classify, and counter malicious drones or birds.

In order to detect and classify a target, its electromagnetic signatures are important. 
One of them is radar cross section (RCS), which is electromagnetic reflectivity due to its geometrical shape and composed material \cite{KnottRCS}.
Since RCS needs to satisfy a far-field condition, the term “reflectivity” is used in this paper for both near and far-field conditions.
The other signature is micro-Doppler, which refers to the small fluctuations in the Doppler shift caused by the target's local moving parts, e.g., the rotating blades of drones, or the flapping wings of birds.
Since the illuminator to the target and the sensing device are at different locations, the bistatic configuration is particularly interested.

\begin{figure}[t]
    \centering
    \includegraphics[width=3.312in, height=2.0in]{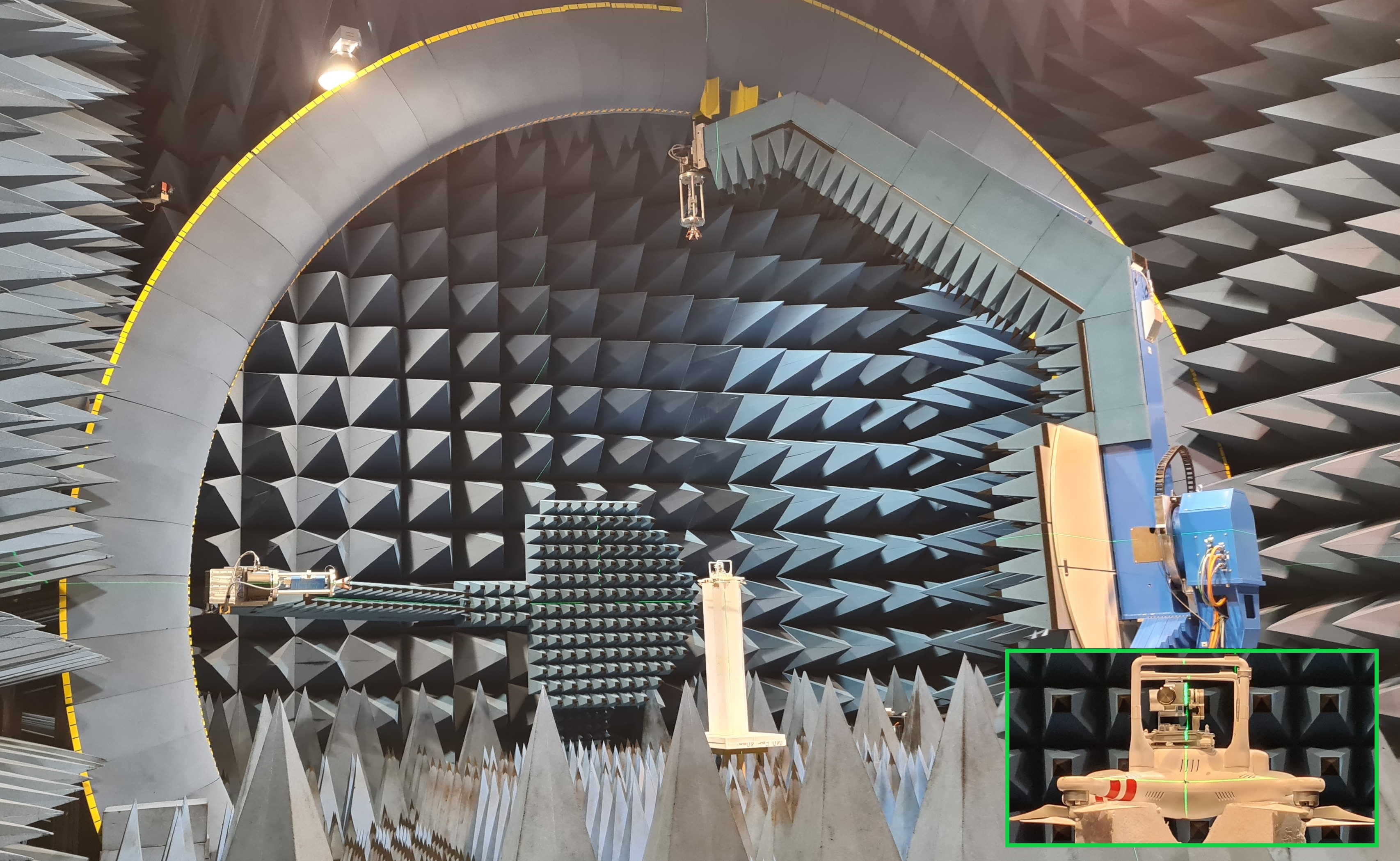}
    \caption{Reflectivity measurement in BiRa: the two gantries can scan the object at the center of the turntable more than the dimension of the hemisphere for both incident and scattering. Note: The arch antennas are not part of BiRa.}
    \label{fig_BiRa}
\end{figure}

Therefore, this paper presents measurement and analysis of the drone bistatic reflectivity and micro-Doppler.
The measurement is carried out by a state-of-the-art measurement system, BiRa at Ilmenau University of Technology. 
This measurement facility is designed particularly for bistatic reflectivity and micro-Doppler measurement, but not limited to, because it can be extended according to the applications \cite{Thoma2}.  
It comprises two pivoting gantries, each equipped with a transmitting (Tx) and receiving (Rx) antenna, respectively. 
The target object of interest is positioned at the center of the turntable as in \autoref{fig_BiRa}.
Therefore, the target can be illuminated and observed from any desired direction within the upper half space.

In section II, the target reflectivity measurement and analysis are described.
Section III discusses the bistatic micro-doppler signature and the paper ends with Section IV conclusion and outlook.

%% file: 2Reflectivityv2.tex
\subsection{Radar Cross Section and Reflectivity}
The well-known definition is "\textit{The Radar Cross Section (RCS) of a target is the projected area of a metal sphere that would scatter the same power in the same direction that the target does}."\cite{KnottRCS}

In general, RCS $\sigma$ can be described as follows:
\begin{equation}       
    \sigma=\lim_{d_{Tx},d_{Rx}\rightarrow\infty}4\pi\cdot d_{Rx}^{2}\cdot\frac{\vert E_{scat}\vert ^{2}}{\vert E_{inc}\vert ^{2}}. 
    \label{equ_bi_RCS}
\end{equation}
Here, $E_{inc}$ is the electric field strength of the incident wave impinging on the target, therefore it is independent of the propagation loss and distance $d_{Tx}$ between the transmitter and the target. 
$E_{inc}$ is the electric field strength of the scattered wave at the distance $d_{Rx}$ from the target \cite{KnottRCS}.

Without the limit on distances $d_{Tx}$ and $d_{Rx}$ in \eqref{equ_bi_RCS}, RCS is a function of distance (or range). 
This distance dependency can be eliminated when the incident wave is a plane wave and the scattered power is measured at infinity \cite{IEEEDictionary}.
However, it should not be interpreted that the radar must be placed at an infinite distance from the target in the RCS measurement, instead far-field distance is applied.
Therefore, \eqref{equ_bi_RCS} is valid only under far-field conditions.

Since the far-field distance is ${2\cdot D^2}/\lambda$, the electrically larger targets have large far-field distances.
As an example, the scattering far-field distance of a 4 m-long vehicle at 5.9 GHz can be over 600 m. 
Normally, the vehicular communication and detection are within 250 m \cite{Li}.
The RCS measurement in the anechoic chamber, like \autoref{fig_BiRa} with 6 m diameter, can not fulfill the far-field condition.

Electromagnetic scattering behaves differently in far-field and near-field. Therefore, the term "reflectivity, $\mathscr{R}$" is used, as in this paper \cite{Schwind}, to generalize the far-field and near-field conditions.
\begin{equation} 
   \mathscr{R}=\frac{\vert E_{scat}\vert ^{2}}{\vert E_{inc}\vert ^{2}}
   \label{equ_reflectivity}
\end{equation}

\begin{table}[t]
    \caption{REFLECTIVITY MEASUREMENT SETUP}
    \centering
    \begin{tabular}{|l|l|}
    \cline{1-2}
        System                          & Vector Network Analyzer (VNA)    \\ \cline{1-2}
        Frequency sweep                 & 2.0 GHz-18.0 GHz, 1601 steps     \\ \cline{1-2}
        \multirow{2}{*}{Angular sweep}  & Gantries: 10:5:180               \\ \cline{2-2}
                                        & Turn-Table: 0 and 45             \\ \cline{1-2}
        Polarizations                   & Full polarizations               \\ \cline{1-2}
        Target                          & DJI Phantom 2                           \\ \cline{1-2}
    \end{tabular}
    \label{tab_reflectivity_measurement_setup}
\end{table}

\subsection{Scenario of Interest}
\begin{figure}[b]
    \centering
    \includegraphics[width=0.95\columnwidth]{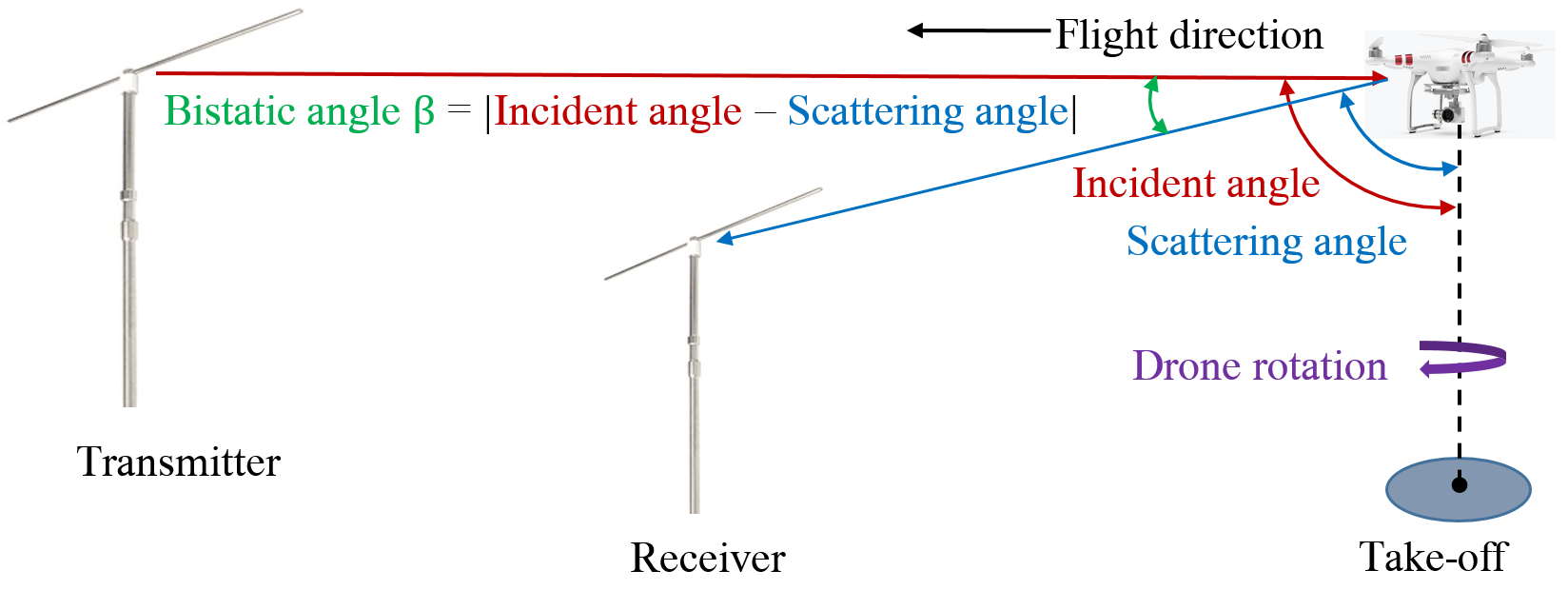}
    \caption{Analyzed scenario for drone detection}
    \label{fig_Focus_scenario}
\end{figure}

The selected scenario for analysis in this paper is illustrated in \autoref{fig_Focus_scenario}. 
It is a flyover scenario, and the assumption is that the transmitter, drone, and receiver lie in a straight line.
The drone will fly over the receiver until the position between the transmitter and the receiver, which introduces bistatic and forward scattering cases
Therefore, the reflection comes from the lower hemisphere of the drone with bistatic angles $\beta$ of 10° to 180°.
For simplicity, take-off and landing are omitted.

\subsection{Bistatic Reflectivity Measurement in BiRa}
The mentioned scenario can be transformed into the angular sweeps of the two gantries as in \autoref{tab_reflectivity_measurement_setup}.
To measure reflection from the lower hemisphere of the drone, it is placed upside down on the styrofoam support at the center of the turntable, in \autoref{fig_BiRa}.
Tx and Rx antennae are RFSpin QRH20E, VNA is from Keysight, and the transmitted signal is the stepped sine signal.
For full polarization measurement, the VNA measures 4 times for each tx-target-rx constellation.
The number of frequency sweeps is 1601 from 2 GHz to 18 GHz. 

The measurement is performed with and without the drone.
The measurement with the drone is named $S_{21, DUT\&BG}$, which is also known as the forward transmission coefficient with device-under-test (DUT) and background (BG). 
The second is a background measurement and is named $S_{21, BG}$.
Here $S_{21}$ is a function of different bistatic angles $\beta$, which is a difference of the incident and scattering angles, and measured frequency bands $f$.
   
\subsection{Reflectivity Post-Processing}
\begin{figure}[b]
    \centering
     \includegraphics[width=0.9\columnwidth]{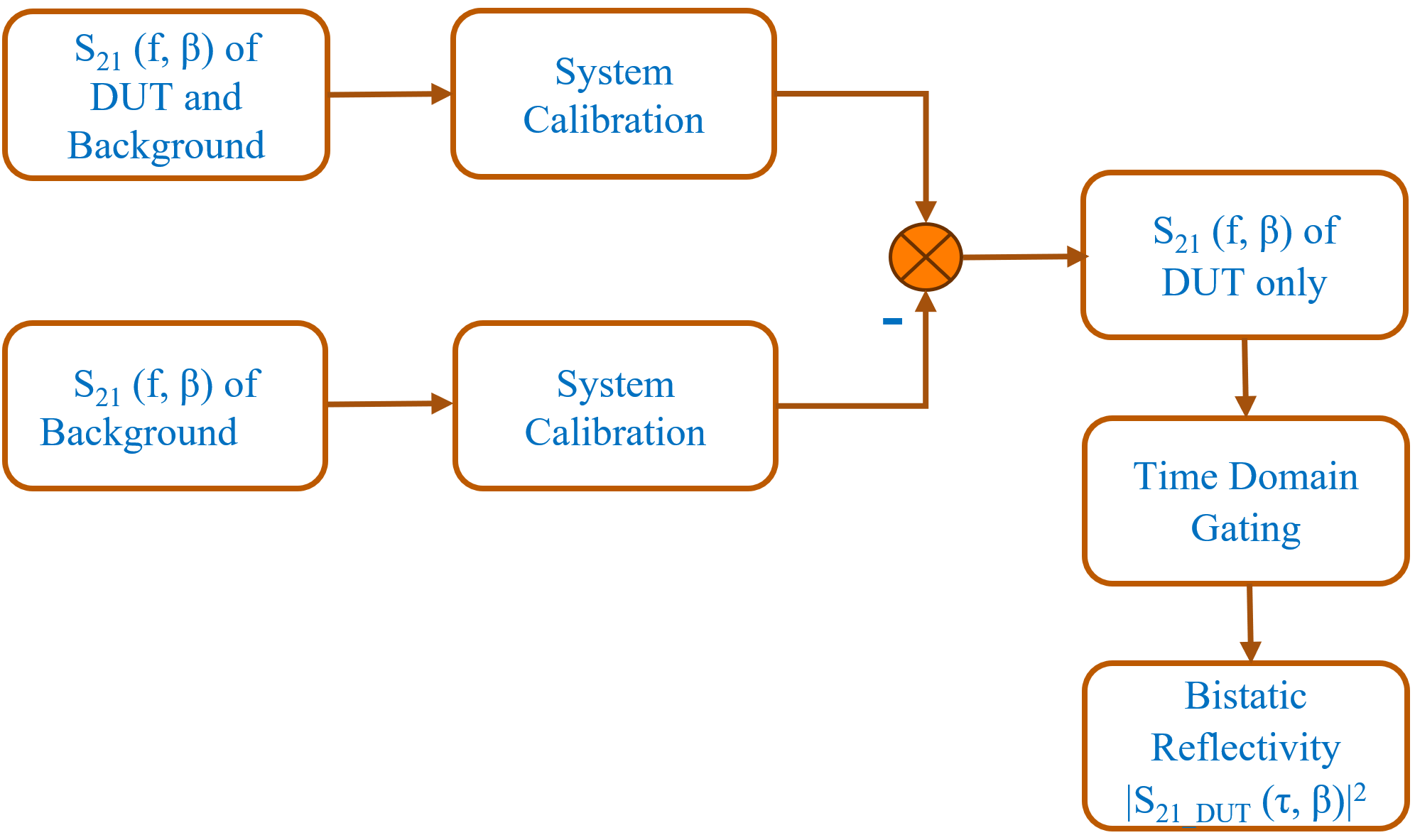}
    \caption{The block diagram of signal processing}
    \label{fig_reflectivity_signal_processing}
\end{figure}

There are three main processes: (1) system calibration, (2) background subtraction, and (3) time domain gating, as illustrated in \autoref{fig_reflectivity_signal_processing}.
In system calibration, the measured data are de-embedded from antenna frequency response and VNA through measurement with an attenuator, which is used to avoid saturation at the receiver.
The strong reflections of objects within the anechoic chamber are eliminated by the background subtraction.
The above-mentioned processes are done in the frequency domain.
Furthermore, the residual reflections are filtered out by time domain gating in the time domain.
Then, the processed data are normalized by maximum reflectivity across bistatic angles, frequency bands, and polarizations.
Since the measurement is in the near-field, the propagation loss can not be compensated by the Friis equation.
Therefore, the presented reflectivity is for a illumination from 3 m distance and scattering at a 3 m distance from the drone.

\begin{figure}
    \centering
    \subfloat[Turn table at 0°, angular sweep 10° to 180° and polarization HH]{\includegraphics[width=0.95\columnwidth]{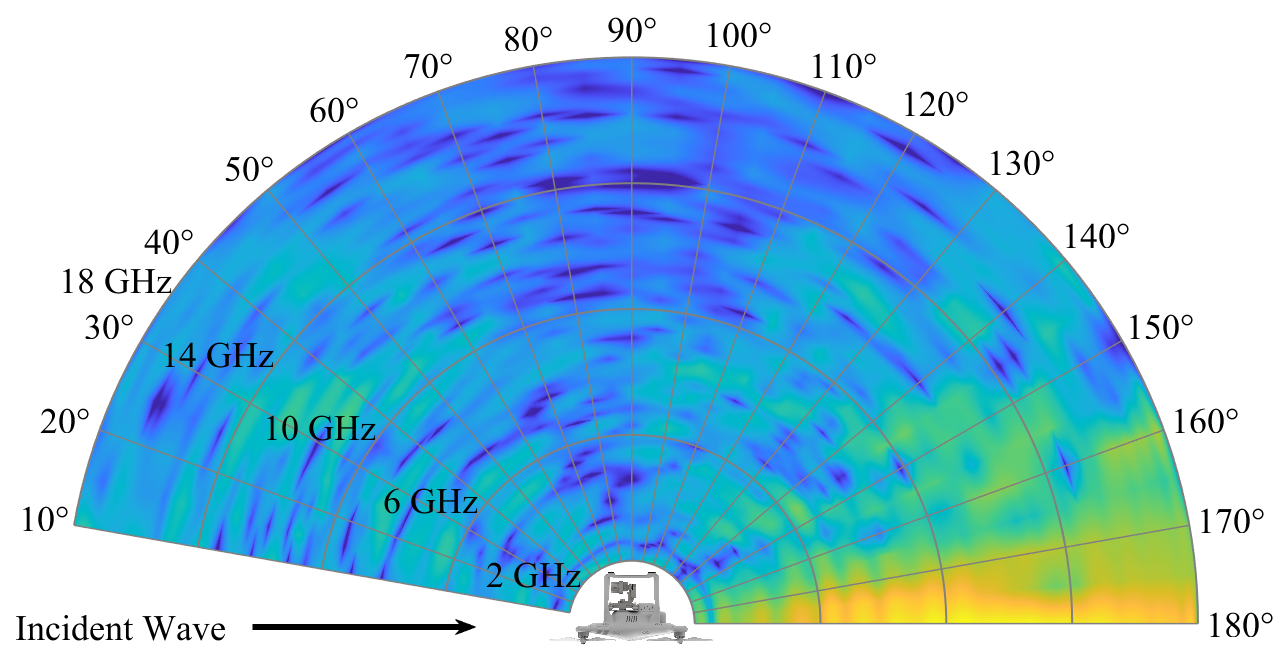}%
    \label{fig_BiRa_0_HH}} \\

    \subfloat[Turn table at 0°, angular sweep 10° to 180° and polarization HV]{\includegraphics[width=0.98\columnwidth]{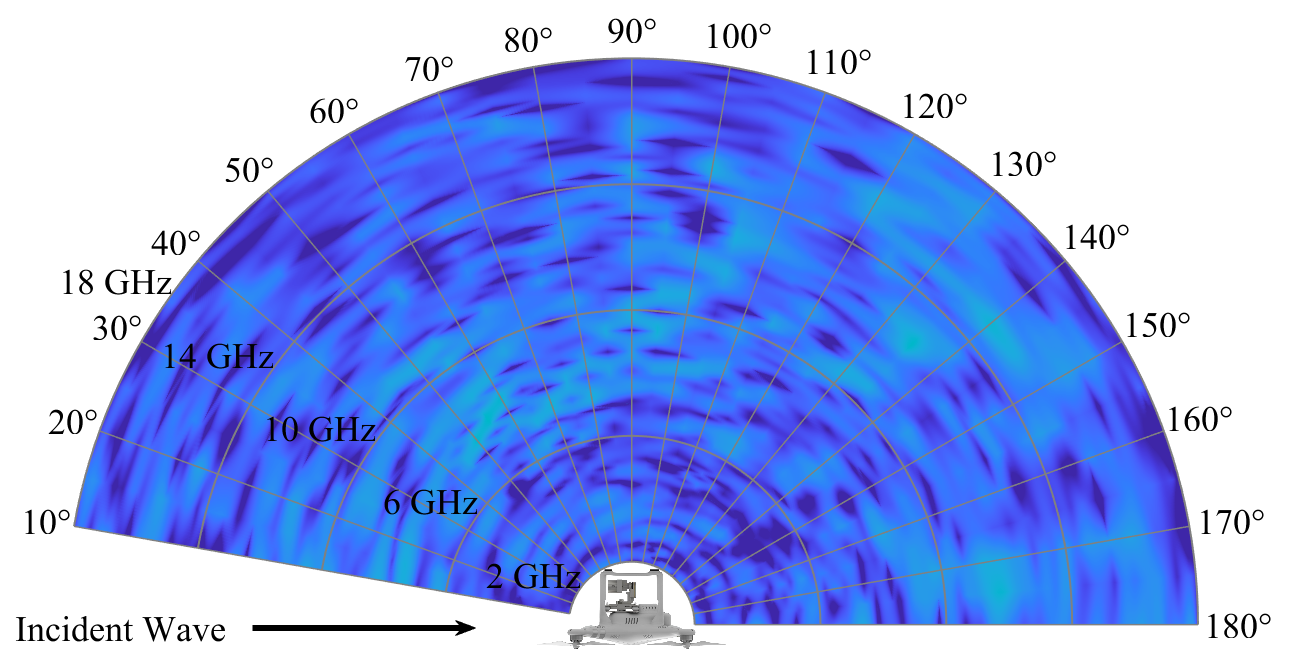}%
    \label{fig_BiRa_0_HV}}\\

    \subfloat[Turn table at 45°, angular sweep 10° to 180° and polarization HH]{\includegraphics[width=0.98\columnwidth]{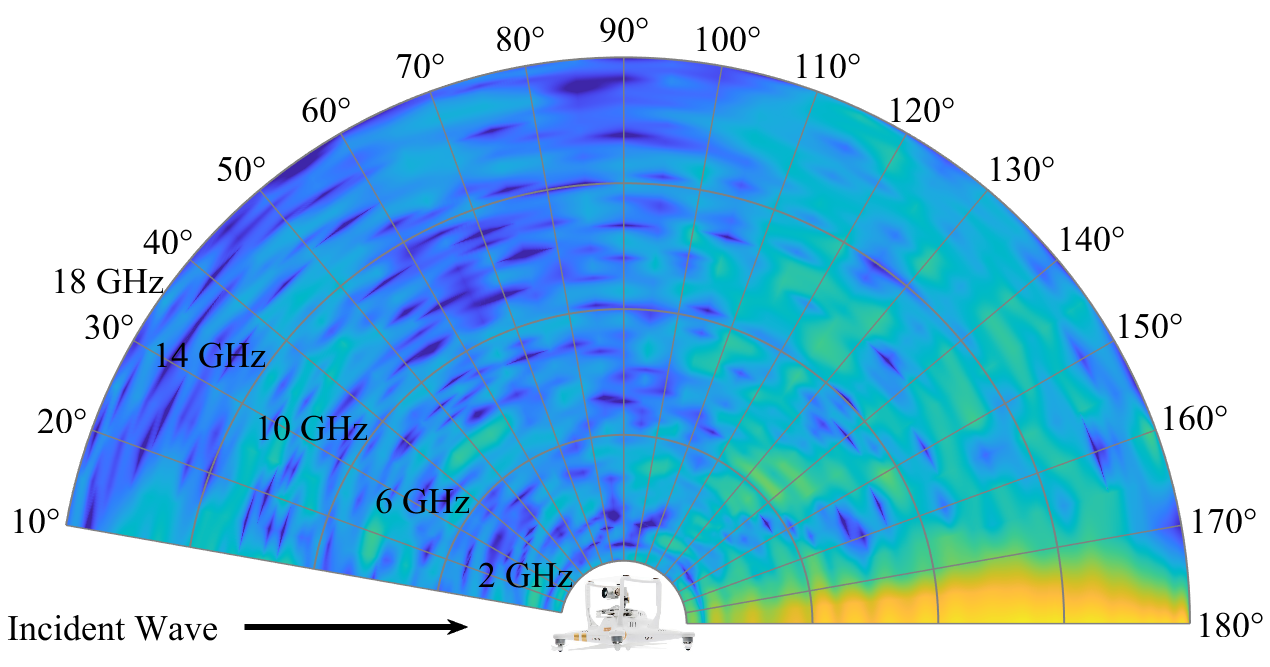}%
    \label{fig_BiRa_45_HH}}\\
    
    \subfloat[Simulated reflectivity by FEKO, using \ref{fig_BiRa_0_HH} setting]{\includegraphics[width=0.98\columnwidth]{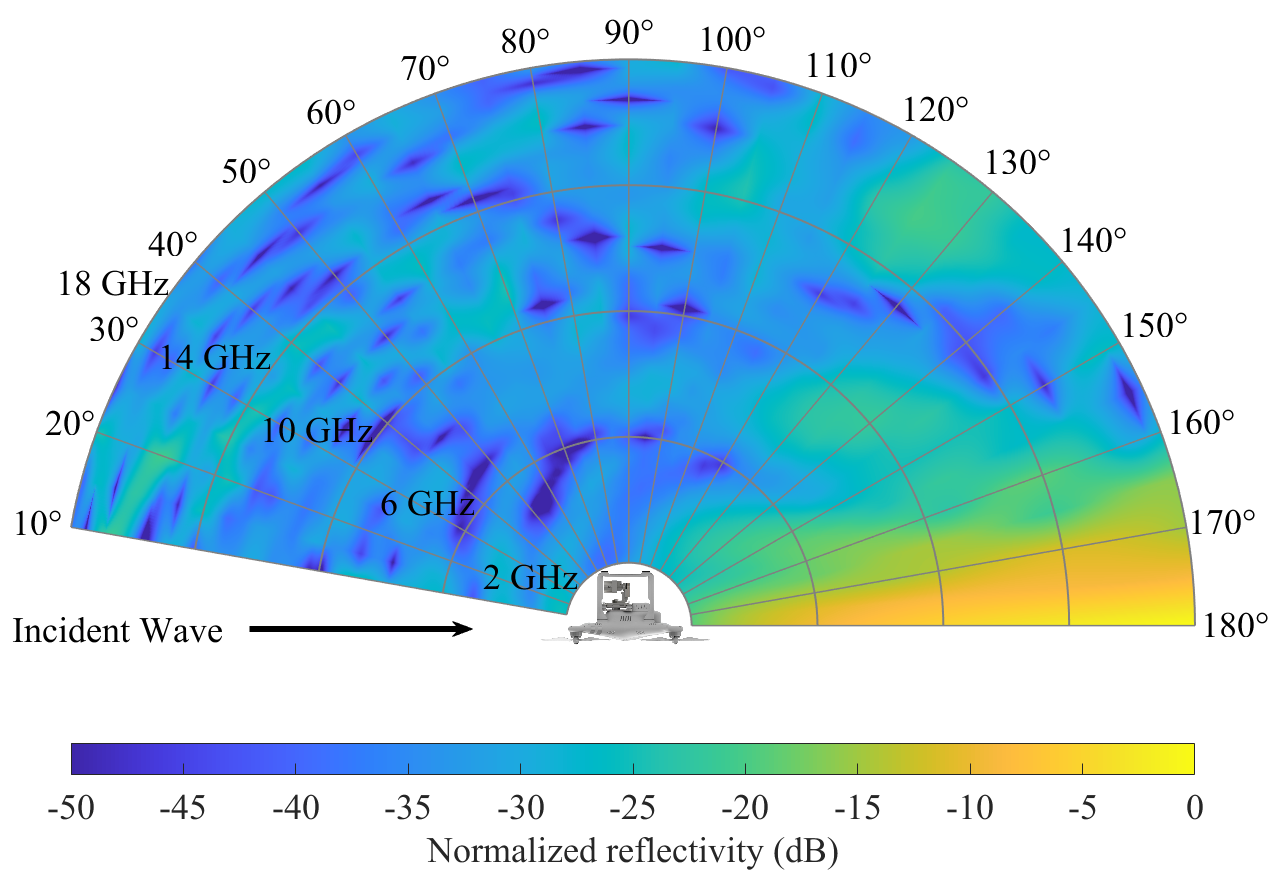}%
    \label{fig_FEKO_0_HH}}
\caption{Normalized reflectivity of DJI P2 for different bistatic viewing angles (10° to 180°) and polarizations over frequency of 2 GHz to 10 GHz.}
\label{fig_result}
\end{figure}

\subsection{Analysis of the Results}
The normalized bistatic reflectivity of DJI Phantom 2 over different bistatic angles $\beta$ and frequency bands $f$ are presented in \autoref{fig_result}.
As expected, the bistatic reflectivity varies according to the bistatic angles and the operating frequency. 
The measured reflectivity in BiRa, \autoref{fig_BiRa_0_HH}, is compared with FEKO full-wave simulation, \autoref{fig_FEKO_0_HH}, and it can be seen that both are well agreed by visual inspection. 

Due to the dependency of reflectivity on polarization, in \autoref{fig_BiRa_0_HH} and \autoref{fig_BiRa_0_HV}, acquiring the target reflectivity in full polarization could help for the further target classification task, especially for data demanding algorithms.

In \autoref{fig_BiRa_0_HH}, the drone is illuminated between its two arms, and in \autoref{fig_BiRa_45_HH}, it is illuminated at its arm. 
This 45° rotation of the drone on its axis (or illumination on the different geometrical parts of the target) can introduce a different reflectivity signature.

\subsection{Further Use of the Measured Data}
The acquired data are the electromagnetic signatures of the target in different polarizations, bistatic angles, and frequency bands. 
Therefore, it is of great interest to analyze the radar performance in target detection and classification. 
For the current research project 6G-ICAS4Mobility, the data will be applied to train machine learning algorithms for target detection and classification.

%% file: 3Micro-Doppler.tex
\begin{figure}[t]
\centering
    \includegraphics[width=0.45\textwidth, height=1.8in]{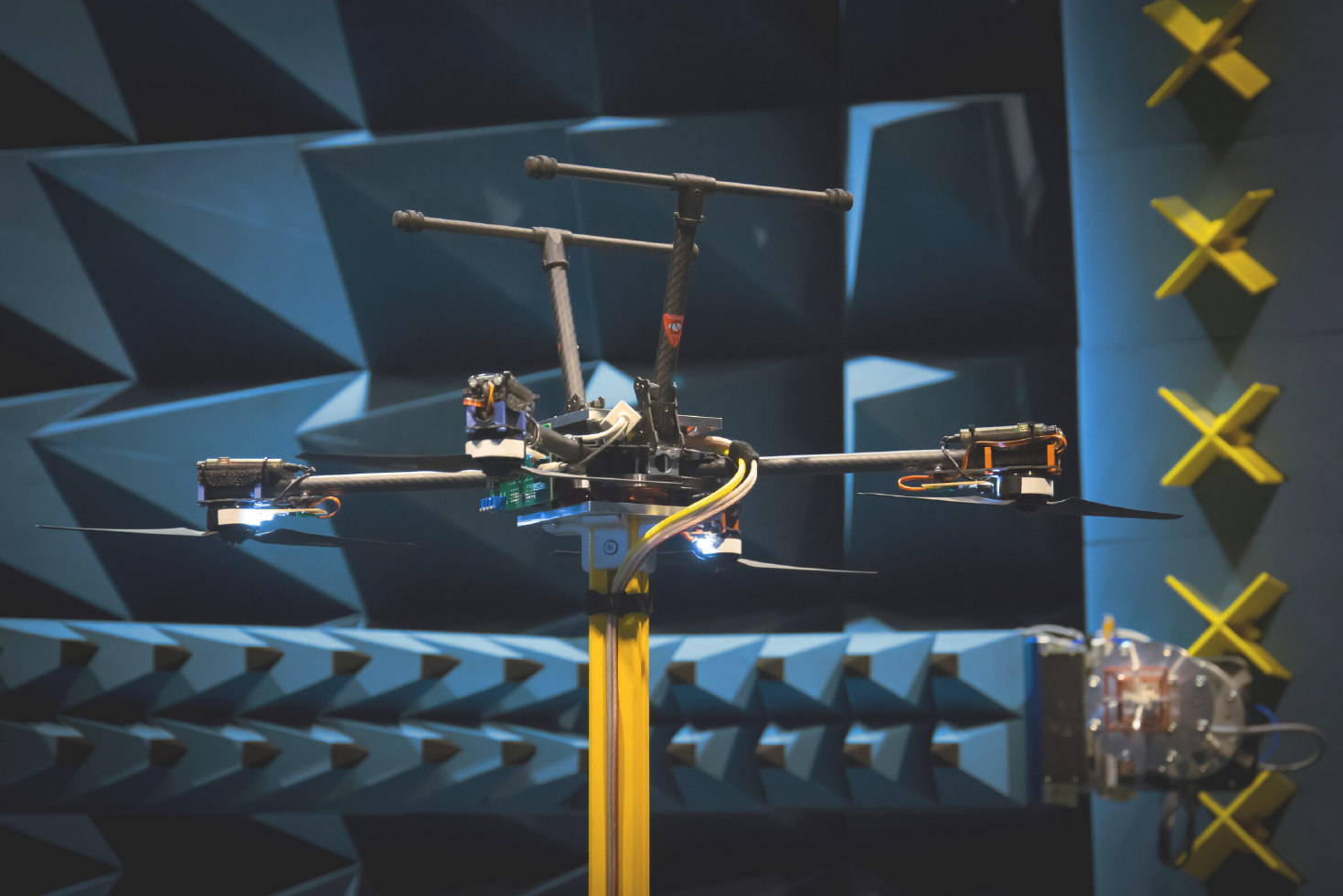}
    \caption{Micro-Doppler measurement in BiRa.}
    \label{fig_mockup}
\end{figure}

On the dynamic reflectivity side, micro-Doppler refers to the modulation on the radar returns due to Doppler contributions caused by the local movements of the target. 
Therefore, micro-Doppler provides valuable information about the target's structure and motion, which can be used to identify the type of target and distinguish it from other objects.

Bistatic geometry and orthogonal frequency-division multiplexing (OFDM) waveform for communication are the most common situations in ICAS.
Thus, it is important to understand how micro-Doppler appears under these conditions. 

\subsection{Micro-Doppler under OFDM-based waveforms}
About OFDM-based waveforms, they reflect the target's micro-movements in a way that differs from typical continuous wave (CW) and linear frequency modulation (LFM) radars, since each frequency component of the OFDM signal produces a different Doppler response. 
This effect can be neglected in narrowband systems, but it is important in broadband environments, and even more important for ultra-wideband (UWB) and high-frequency sensors. 
Besides, the use of multiple subcarriers increases system performance with respect to the SNR of the micro-Doppler signature \cite{Sen2014}.

\subsection{Micro-Doppler under Bistatic Configuration}
Regarding bistatic micro-Doppler, measurement and analysis from various aspect angles are important tools for characterizing targets, since they can provide additional information about the target's motion and structure, which is not available from monostatic measurements alone \cite{smith2008radar}. 

The development of advanced signal processing techniques for bistatic micro-Doppler measurement can contribute greatly to solve target classification problems under ICAS.

In order to develop these techniques, although some advances can be made with simulation tools, it is important to have access to a large amount of real target measurement data. 

However, despite the importance of bistatic micro-Doppler measurements in ICAS systems, there is currently a limited amount of data and research on this topic compared to monostatic micro-Doppler. 
This is partly due to the technical challenges associated with bistatic measurement setups. 

BiRa can play an interesting role in the way to solve this problem since it is able to perform multi-aspect bistatic micro-Doppler measurements in an automatized way.
\begin{table}[t]
    \caption{MICRO-DOPPLER MEASUREMENT SETUP}
    \centering
    \begin{tabular}{|c|c|c|}
        \hline
        \multirow{3}{4em}{\textbf{System}} & Waveform & OFDM-like \\
        \cline{2-3}
        & Central frequency & 3.7 GHz \\
        \cline{2-3}
        & Bandwidth & 200 MHz \\
        \hline
        \multirow{3}{4em}{\textbf{OFDM Settings}} & Total number of carriers & 1600 \\
        \cline{2-3}
        & Carriers with energy & 1280 \\
        \cline{2-3}
        & Symbol duration & $8 \mu s$ \\
        \hline
        \multirow{5}{4em}{\textbf{Drone }} & Number of propellers & 4 \\
        \cline{2-3}
        & Blades per propeller & 2 \\
        \cline{2-3}
        & Propeller radius & 16.55 cm \\
        \cline{2-3}
        & Rotation speed & 25 Hz \\
        \cline{2-3}
        & Propeller material & carbon fiber \\
        \hline
    \end{tabular}
    \label{tab_MicroDoppler_measurement_setup}
\end{table}

\subsection{Bistatic Micro-Doppler Measurement in BiRa} 
\label{MicroDoppler_BiRa_measurements}

\autoref{tab_MicroDoppler_measurement_setup} details the measurement setup for this work.
A wideband OFDM-based transmit signal, called Newman sequence with constant spectral magnitude and minimal crest-factor \cite{Boyd1986}, was used.
Additionally, many bistatic angle combinations were employed, from which six configurations will be analyzed in this paper.

A custom drone target system was developed for this measurement. This target allows remote setting of the propeller's angular velocities, which can not be achievable with standard commercial drones. 
This is especially important for micro-Doppler analysis purposes since it is desirable to have ground truth of the propellers' speeds.
This system, presented on \autoref{fig_mockup}, is composed of a Tarot IRON MAN 650 mechanical structure, four DJI E800 motors with Electronic Speed Controls (ESC), four custom-built optical speed sensors, an Arduino Nano programmable board with Ethernet shield, and a firmware for controlling the rotors, including a closed control loop which ensures that the desired speed is applied and kept constant.

\begin{figure}[b]
    \centering
    \includegraphics[width=0.49\textwidth, height=1.5in]{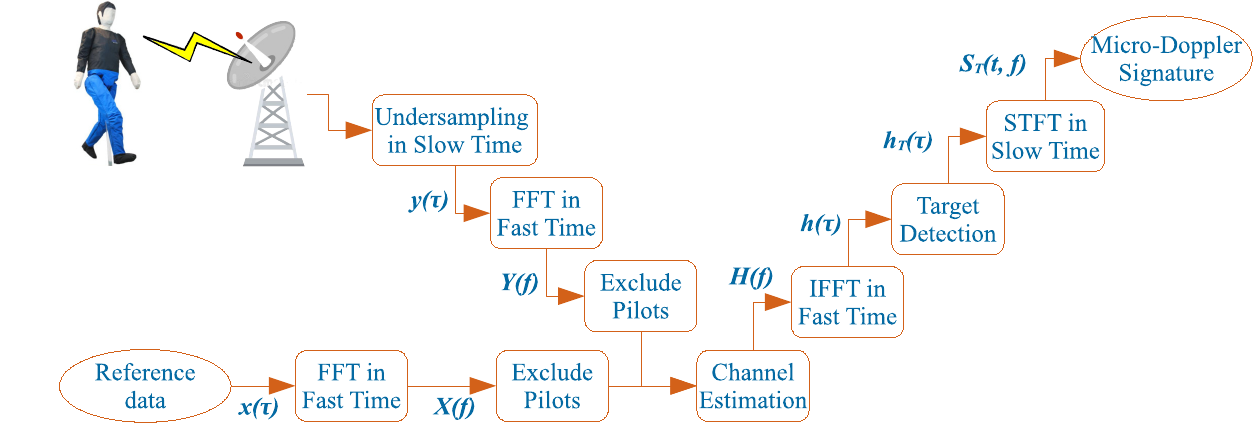}
    \caption{The block diagram of OFDM micro-Doppler signal post-processing}
    \label{fig_signal_processing}
\end{figure}

\subsection{OFDM Micro-Doppler Post-Processing} \label{signal_processing}

For micro-Doppler analysis, the signal was processed in slow-time, i.e., a vector of the returns of the target in a single range bin, along different symbols. 

To achieve this, the signal processing scheme shown in \autoref{fig_signal_processing} is applied. First, both the received signal and the reference from time domain ($y(\tau)$ and $x(\tau)$) are converted into frequency domain ($X(f)$ and $Y(f)$), respectively. Then, the pilot frequencies are excluded, and the channel is estimated by doing $H(f) = \frac{Y(f)}{X(f)}$ \cite{Braun2014}.

Finally, this process is followed by target detection and Spectrogram calculation of the slow-time profile, in order to generate a time-frequency representation.

Since high symbol repetition frequency is used in this measurement, a subsampling factor of 8 is applied to the data before processing in order to reduce both the amount of data and the slow-time sampling frequency.
Then, lower processing cost and better visualization of the data by focusing the analysis to the interest bandwidth are achieved.

\subsection{Analysis of the Results} \label{first_results}
\begin{figure}[t]
    \centering
    \includegraphics[width=0.45\textwidth, height=2in]{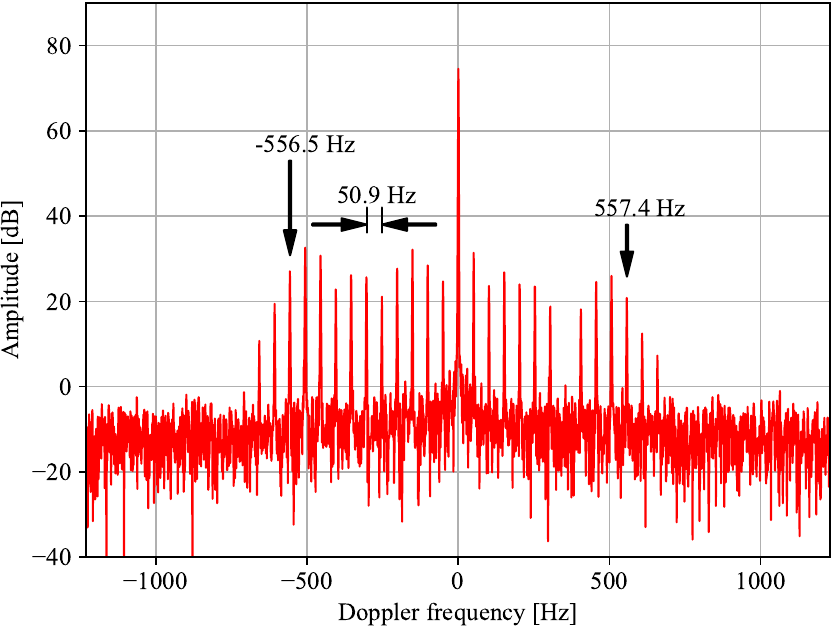}
    \caption{Rotating propeller micro-Doppler signature in frequency domain (32768 symbols).}
    \label{fig_FFT_copter}
\end{figure}

A micro-Doppler frequency signature of one rotating propeller is presented in \autoref{fig_FFT_copter}. 
There, a strong impulse centered in $f = 0$ Hz and many impulses on both sides of this stronger one can be seen above the noisy basis. 
The central impulse occurs due to the static parts of the drone, while the other impulses come from the rotating propeller. 

The first important observation is that these impulses are separated by $\Delta f = N_b f_{rot}$, where $f_{rot}$ is the propeller rotation frequency and $N_b$ is the number of blades in the propeller.
This can be explained by the fact that the returns of a rotating propeller are periodic in time domain. 
The Discrete-Time Fourier transform (DTFT) of any signal $x[n]$, with period $N$ and $\omega_0 = \frac{2\pi}{N}$, is always a sequence of impulses separated by the period, which can be described as
\begin{equation}
    X(j\omega) = \frac{2\pi}{N} \sum_{k=-\infty}^{\infty}{[a_k \cdot \delta(\omega - \frac{2\pi k}{N})]},
    \label{equ_DFT}
\end{equation}
where $a_k$ denotes the coefficients of the Fourier series of the repeating part of the signal:
\begin{equation}
    a_k = \sum_{n=0}^{N-1}{x[n] e^{-j 2\pi \frac{k}{N}n}}.
    \label{equ_coef_DFT}
\end{equation}

Therefore, the distance among these impulses is directly related to the rotation speed and the number of propellers.
In this case, the rotation frequency was $f_{rot} = 1500 \text{ rpm} = 25$ Hz and $N = 2$, so $\Delta f = 50$ Hz, which is consistent with what can be seen in \autoref{fig_FFT_copter}. 

\begin{figure}[t]
    \centering
    \includegraphics[width=0.45\textwidth, height=2in]{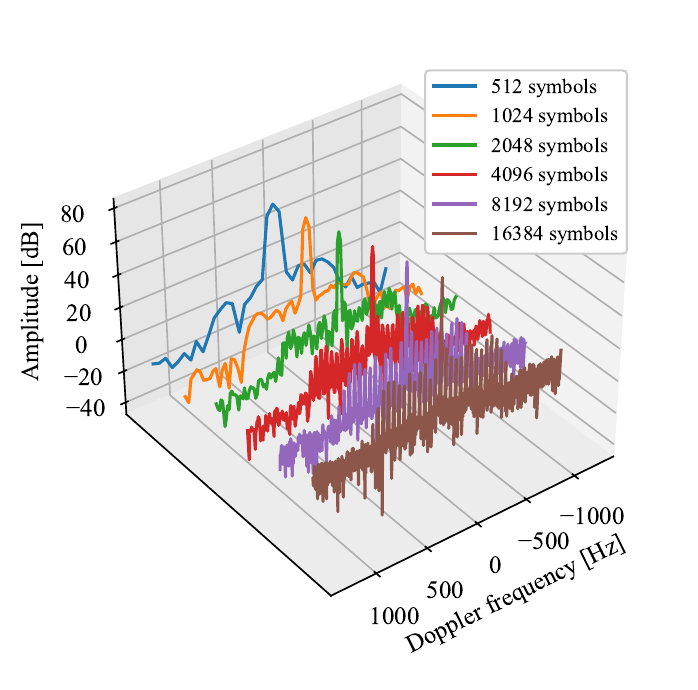}
    \caption{Comparison of micro-Doppler signatures with fewer samples available. The legend indicates the number of symbols used in each measurement. The sampling frequency is the same in all cases.}
\label{fig_FFT_copter_undersampled}
\end{figure}

In order to be able to clearly identify the frequency domain impulses, the measurement must have a long observation time, which implies a high number of samples. 
With fewer samples, i.e short observation time, the signal gets undersampled in frequency domain, making it impossible to differentiate the impulses inside the signal. 
This effect can be seen in \autoref{fig_FFT_copter_undersampled}.

A second important feature that can be seen, is the micro-Doppler spread which will be now analyzed.

In bistatic geometries, the Doppler frequency of a point target is given by \cite{willis2005bistatic}:

\begin{equation}
    f_D = \frac{-2\| \bm{v} \|\cos{\frac{\beta}{2}}\cos{\phi}\sin{\theta}}{\lambda},
    \label{equ_fd}
\end{equation}
where $\beta$ is the bistatic angle and $\phi$ and $\theta$ are the azimuth and elevation angles between the bistatic bisector and the target speed $\bm{v}$.
By exploiting \eqref{equ_fd}, the micro-Doppler spread of rotating propellers in bistatic geometries can be inferred. The translational movement of the target does not contribute to the micro-Doppler spread, so it will be ignored in this analysis.

A propeller can be regarded as a continuous set of point targets, rotating with angular velocity $\omega$ and radius $l \in [0, L]$. If the blade length $L$ is much smaller than the range, the Doppler frequency of each propeller point can be written as

\begin{equation}
    f_D = \frac{-2 \omega l \cos{\frac{\beta}{2}}\cos{(\phi_0 + \omega t)}\sin{\theta}}{\lambda}.
    \label{equ_fd_prop_point}
\end{equation}

The micro-Doppler spread can then be written as

\begin{equation}
    B_D = 2{f_D}_{max} = \frac{4 \omega L \cos{\frac{\beta}{2}}\sin{\theta}}{\lambda},
    \label{equ_doppler_spread}
\end{equation}

Therefore, the micro-Doppler spread depends on the bistatic angle $\beta$. 
\autoref{fig_bistatic_angles} shows from measurement data how the micro-Doppler spread decreases when the bistatic angle increases.
Therefore, we have maximum micro-Doppler spread in the monostatic case ($\beta = 0^{\circ}$) and minimum micro-Doppler spread in forward scattering configuration ($\beta = 180^{\circ}$). 
It can also be seen, that distance among impulses does not change with respect to bistatic angles, since they depend only on the rotation frequency and the number of blades.

In \autoref{fig_FFT_copter}, the maximum velocity is given by $\|\bm{v}\| = 2 \pi f_{rot} L = 2\pi \times25\times0.1655 = 26$ m/s, where $L = 0.1655$ m.
As $\beta = 60^{\circ}$ and $\theta = 90^{\circ}$, then we can calculate a maximum Doppler frequency ${f_D}_{max}$= 555.4 Hz, which is very similar to what we can see in the image, from measurement data.

In order to be able to clearly identify the micro-Doppler spread, the measurement must have a sampling frequency higher than the maximum Doppler from the target, which avoids aliasing.

\subsection{Further Use of the Measured Data}

There is still a vast range of investigations that must be performed, such as high-range resolution micro-Doppler, usage of other time-frequency representations to achieve a better frequency resolution $\times$ time resolution balance, analysis of the forward scatter case and its particularities, and performing joint micro-Doppler and static reflectivity drone classification in bistatic and multistatic distributed configurations.

%% file: 4Conclusions.tex
This paper presents a new look into drones' reflectivity and micro-Doppler signatures under the context of ICAS, which are bistatic geometry (transmitter-target-receiver), communication frequency bands, and OFDM waveform.
These three aspects are different from classic radar systems.
The BiRa measurement system offers new capabilities, which allow us to measure the reflectivity and micro-Doppler in the above-mentioned ICAS context.

Therefore, we measured and analyzed the reflectivity in multiple aspects of scanning with many communication frequency bands in full polarization.
Results show, by comparing with the full-wave simulation, that the measured reflectivity is good for further processes, eg. modeling.

We also use BiRa measurements to check how the micro-Doppler signatures of rotating propellers behave in the referred context. Results show that, in frequency domain, periodic impulses appear, whose period is related to the propeller rotation speed and the number of blades. 
Another effect analyzed in the results is how micro-Doppler spread decreases when the bistatic angle increases, being maximum for the monostatic geometry and minimum for the forward scattering case.

The high-range resolution (HRR) analysis, providing extended target reflectivity and micro-Doppler, will be presented and discussed in upcoming publications.

\begin{figure}[t]
    \centering
    \includegraphics[width=0.48\textwidth, height=2in]{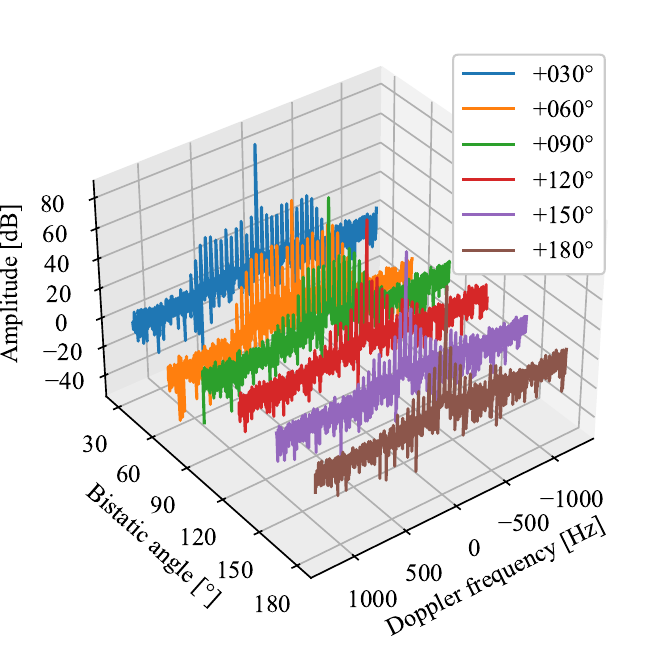}
    \caption{Comparison of frequency-domain micro-Doppler signatures of a rotating propeller with different bistatic angles (16384 symbols). The legend indicates the bistatic angle in each measurement.}
\label{fig_bistatic_angles}
\end{figure}